\documentclass[prd,aps,showpacs,tightline,twocolumn,nofootinbib,preprintnumbers]{revtex4}
\usepackage{graphicx}
\usepackage{amssymb}
\usepackage{amsmath}

\preprint{IPMU-10-0145}
\preprint{KEK-TH-1394}

\begin{document}

\newcommand{\beq}{\begin{equation}}   
\newcommand{\eeq}{\end{equation}}
\newcommand{\bea}{\begin{eqnarray}}   
\newcommand{\eea}{\end{eqnarray}}
\newcommand{\bear}{\begin{array}}  
\newcommand {\eear}{\end{array}}
\newcommand{\bef}{\begin{figure}}  
\newcommand {\eef}{\end{figure}}
\newcommand{\bec}{\begin{center}}  
\newcommand {\eec}{\end{center}}
\newcommand{\non}{\nonumber}  
\newcommand {\eqn}[1]{\beq {#1}\eeq}
\newcommand{\la}{\left\langle}  
\newcommand{\ra}{\right\rangle}
\newcommand{\ds}{\displaystyle}
\def\SEC#1{Sec.~\ref{#1}}
\def\FIG#1{Fig.~\ref{#1}}
\def\EQ#1{Eq.~(\ref{#1})}
\def\EQS#1{Eqs.~(\ref{#1})}
\def\GEV#1{10^{#1}{\rm\,GeV}}
\def\MEV#1{10^{#1}{\rm\,MeV}}
\def\KEV#1{10^{#1}{\rm\,keV}}
\def\lrf#1#2{ \left(\frac{#1}{#2}\right)}
\def\lrfp#1#2#3{ \left(\frac{#1}{#2} \right)^{#3}}
\newcommand{\phih}{\hat{\phi}}
\newcommand{\phit}{\tilde{\phi}}

%

\title{
Higgs Chaotic Inflation in Standard Model and NMSSM
}

\author{
Kazunori Nakayama${}^{(a)}$ and
Fuminobu Takahashi${}^{(b)}$
}

\affiliation{
${}^{(a)}$ Theory Center, KEK, 1-1 Oho, Tsukuba, Ibaraki 305-0801, Japan\\
${}^{(b)}$ Institute for the Physics and Mathematics of the Universe,
University of Tokyo, Chiba 277-8583, Japan
}

\date{\today}

\begin{abstract}
  We construct a chaotic inflation model in which the Higgs fields play the role of the inflaton in the 
  standard model as well as in the singlet extension of the supersymmetric standard model.   The key idea is to allow
  a non-canonical kinetic term for the Higgs field. The model is a realization of the recently proposed running kinetic inflation, in which
  the coefficient of the kinetic term grows as the inflaton field.   The inflaton potential
  depends on the structure of the Higgs kinetic term. In the simplest cases, the inflaton potential is
   proportional to $\phi^2$ and $\phi^{2/3}$  in the standard model and NMSSM, respectively. It is also possible to have a flatter inflaton 
   potential.
  \end{abstract}

\pacs{98.80.Cq}

\maketitle

The inflation is strongly motivated by the recent WMAP results~\cite{Komatsu:2010fb}.
It is a non-trivial task to construct a successful inflation model, partly because the properties of
the inflaton are poorly known. The inflaton may be only weakly coupled to the standard
model (SM) sector. In this case, since the number of cosmological observables are limited,
it might be difficult to pin down the inflation model even with the Planck data~\cite{:2006uk}. 
Alternatively, the inflaton may be
a part of the SM or its extensions~\cite{Murayama:1992ua,Kasuya:2003iv,Allahverdi:2006iq}, in which case we may be able to study the properties
of the inflaton at collider experiments such as the LHC. The latter idea has recently attracted 
much attention since the proposal of the SM Higgs inflation~\cite{Bezrukov:2007ep}. 
In the model of Ref.~\cite{Bezrukov:2007ep},
the flat potential is achieved by introducing a
non-minimal coupling to the gravity~\cite{Salopek:1988qh}
(see also Refs.~\cite{Einhorn:2009bh,Lee:2010hj,Ferrara:2010yw,Ferrara:2010in,Kallosh:2010ug,BenDayan:2010yz} 
for the inflation with non-minimal coupling to gravity in supergravity).
In this letter we pursue another approach to the inflation in the SM and its extensions:
we construct a Higgs chaotic inflation model by allowing a non-canonical kinetic term.
 As we shall see below, the model is a realization of the
running kinetic inflation~\cite{Takahashi:2010ky,Nakayama:2010kt}.

Recently, a new class of inflation models was proposed by one of the
authors (FT)~\cite{Takahashi:2010ky}, in which the kinetic term grows
as the inflaton field, making the effective potential flat~\cite{Dimopoulos:2003iy,Izawa:2007qa}.  This
model naturally fits with a high-scale inflation model such as chaotic
inflation~\cite{Linde:1983gd}, in which the inflaton moves over a
Planck scale or even larger within the last $50$ or $60$
e-foldings~\cite{Lyth:1996im}.  This is because the precise form of
the kinetic term may well change after the inflaton travels such a
long distance.  In some cases, the change could be so rapid, that it
significantly affects the inflaton dynamics.  We named such model as
 running kinetic inflation.  Interestingly, the power of the
inflaton potential generically changes in this class of inflation
models. The phenomenological aspects of the running kinetic inflation was
studied in detail in Ref.~\cite{Nakayama:2010kt}. 

First let us consider the Higgs inflation in the SM. In order to identify the Higgs with the inflaton,
there are two issues. First, if  the potential were valid up to large field values, the chaotic
inflation with a quartic potential would occur. However, the quartic chaotic inflation is strongly
disfavored by observation~\cite{Komatsu:2010fb}. Secondly, in order to satisfy the WMAP normalization,
a quartic coupling must be as small as $O(10^{-13})$ which would result in an unacceptably light Higgs mass.
These issues can be avoided if the potential becomes flatter at large field values. 
There are two ways. One is to introduce a non-minimal coupling to gravity~\cite{Salopek:1988qh} and
the other is to make use of the running kinetic term~\cite{Takahashi:2010ky}. We will focus on the latter possibility in this letter.

The key idea is to add the following interaction,
\beq
\Delta {\cal L} \;=\; \xi |H|^2 |D_\mu H|^2,
\eeq
where $H$ is the Higgs doublet, $\xi$ is a numerical coefficient, and $D_\mu$ denotes a gauge covariant derivative~\footnote{
In Ref.~\cite{Germani:2010gm} a different kind of non-canonical kinetic term was considered. 
}. 
Here and in what follows we adopt the Planck unit, $M_P = 1$.
In the unitary gauge, we can write down the Lagrangian for the Higgs $h$:
\beq
{\cal L}\;=\; \frac{1}{2}\left(1+\xi \frac{h^2}{2} \right) (\partial h)^2-\frac{\lambda_h}{4}(h^2-v^2)^2.
\label{SM}
\eeq
For small $h$, the effect of non-canonical kinetic term is irrelevant, while,
for large $h \gtrsim 1/\sqrt{\xi}$, the kinetic term grows, that is why the name ``running kinetic inflation."
The canonically normalized field in this regime is given by
\beq
{\hat h} \;\approx\; \frac{\sqrt{\xi} h^2}{2 \sqrt{2}},
\eeq
and the effective potential becomes
\beq
V({\hat h}) \;\simeq\; \frac{1}{2} \lrf{4 \lambda_h}{\xi} {\hat h}^2.
\eeq
Thus, the quadratic chaotic inflation occurs. We emphasize here that the potential changes from $h^4$ to ${\hat h}^2$
because of the running kinetic term. A large kinetic term makes the effective potential flatter, and it is straightforward to
obtain a flatter potential by increasing the power of $h$ in the coefficient of the kinetic term.
The WMAP normalization gives
$\lambda_h \simeq 10^{-11} \xi$, and so, if $\xi$ is sufficiently large, $\xi \sim O(10^{10})$,
the quartic coupling $\lambda_h$ can be of $O(0.1)$. 
Such a large coupling is analogous to the non-minimal coupling to gravity in Ref.~\cite{Bezrukov:2007ep}, and it may be obtained by tuning
or some UV dynamics~\cite{Dimopoulos:2003iy,Izawa:2007qa}. 
Note that the inflation takes place for sub-Planckian values of $h$, while the value of ${\hat h}$ exceeds
the Planck scale.

Next we apply the same idea to the Higgs inflation in supergravity. 
As is well known, it is difficult to implement the chaotic inflation  in supergravity
because of the exponential pre-factor $e^K$ in the scalar potential, where $K$ is the K\"ahler potential.  
In order to  construct a chaotic inflation model in supergravity, there must be
flat directions in the field space along which the K\"ahler potential does not grow.
The flat direction can be realized by either symmetry or tuning. 
In the latter case we can assume that a certain interaction in the K\"ahler potential
is enhanced, which results in an approximate flat direction~\cite{Izawa:2007qa, NT}.
Instead, we here adopt the symmetry to ensure the flatness, following the construction in Refs.~\cite{Takahashi:2010ky,Nakayama:2010kt}.
In both cases, the inflation dynamics is essentially the same.
In the pioneering paper by Kawasaki, Yamaguchi and Yanagida~\cite{Kawasaki:2000yn}, 
a shift symmetry on the inflaton, $\phi \rightarrow \phi +  \alpha$ ($\alpha$ is a real transformation parameter), 
was introduced so that the K\"ahler potential depends only
on $(\phi - \phi^\dag)$, not on $(\phi + \phi^\dag)$. In Ref.~\cite{Takahashi:2010ky},
the shift symmetry is generalized to $\phi^n \rightarrow \phi^n + \alpha$, based on the idea
that the form of the kinetic term may change after the inflaton traverses more than the Planck scale.

Let us now  construct a Higgs inflation model in the singlet extension of MSSM.
We introduce a chiral superfield, $\phi$, to represent the gauge invariant $H_u H_d$:
\beq
\phi^2 \;\equiv\; H_u H_d,
\label{huhd}
\eeq
where $H_u$ and $H_d$ are the up- and down-type Higgs superfields. 
In the scalar components, we can express
\beq
H_u \;=\; \left(
\bear{c}
0\\
\phi
\eear
\right),~~H_d = \left( 
\bear{c}
\phi\\
0
\eear
\right).
\eeq
We require that the K\"ahler potential for the Higgs fields is invariant under the following
transformation;
\bea
\phi^2 \;\rightarrow\;\phi^2+ \alpha
\label{sym}
\eea
where $\alpha$ is a real transformation parameter. This corresponds to the above-mentioned shift symmetry with $n=2$. 
We will discuss the case of another value of $n$ later.
The symmetry (\ref{sym}) means that the composite field
${\hat \phi} \sim \phi^2$ transforms under a Nambu-Goldstone like
shift symmetry. 

The K\"ahler potential satisfying the shift symmetry (\ref{sym}) must be a
function of $(\phi^2 - \phi^{\dag 2})$:
\beq
K\;=\;  \sum_{\ell=1} \frac{c_\ell}{\ell} \,(\phi^2 - \phi^{\dag 2})^\ell
\label{Kahler}
\eeq
where $c_\ell$ is a numerical coefficient of $O(1)$ and we normalize
$c_2 \equiv -1$; $c_\ell$ is real (imaginary) for even(odd) $\ell$.
Note that the $|H_u|^2$ and $|H_d|^2$ terms are absent. Instead, the kinetic term
for $\phi$ arises from the terms of $\ell \geq 2$, whose contribution is
proportional to $(\phi^2-\phi^{\dag 2})^{\ell-2}|\phi|^{2}$. 
One can show that $(\phi^2-\phi^{\dag 2})$ remains constant 
along the inflationary trajectory by noting that
$(\phi^2-\phi^{\dag 2})$ appears explicitly in the K\"ahler potential and therefore
acquires a large mass during inflation~\cite{Takahashi:2010ky}.
So, we drop terms with $\ell \geq 3$ because
it does not change the form of the kinetic term.

We can impose a discrete $Z_k$ symmetry that is consistent with the
shift symmetry (\ref{sym}).  Requiring $(\phi^2-\phi^{\dag 2})$, an
invariant under the shift symmetry, be also invariant under the
discrete symmetry up to a phase factor, we find that $k$ must be either $2$ or $4$. 
If $k=4$, the $(\phi^2-\phi^{\dag 2})$ would
flip its sign, and so, $c_\ell$ with any odd $\ell$ should vanish. If
$k=2$, there is no such constraint, since $\phi^2$ itself is invariant under $Z_2$.

In order to have a successful inflation, we introduce explicit symmetry breaking terms
in both the K\"ahler and super-potentials:
\bea
\label{Kand W1}
K&=&\kappa|\phi|^2 - \frac{1}{2}  (\phi^2-\phi^{\dag 2})^2 + |X|^2,\\
W&=&\lambda X \phi^2,
\label{Kand W2}
\eea
where the $\kappa$- and $\lambda$-terms are the symmetry breaking terms, and we assume $\kappa, \lambda \ll 1$.
There could be other symmetry breaking terms, but we assume that they are soft in a sense that the shift symmetry
remains a good symmetry at least up to the inflaton field value of $O(10)$.
Here $X$ is a singlet superfield. $X$ can be stabilized at the origin during and after inflation if we add $- a |X|^4$ in the
K\"ahler potential with $a = O(1)$.
The presence of $X$  not only simplifies the inflaton potential,
but also helps to avoid a situation that the inflaton potential becomes negative due to $-3|W|^2$ in the scalar potential~\cite{Kawasaki:2000yn}.
Here and in what follows we impose $Z_4$ symmetry under which $X$ and $\phi^2$ flip the sign, in order to
suppress dangerous couplings such as $\int d^2 \theta\, X$. The charge assignment of $X$ and $\phi$ are shown in Table~\ref{charge}.

It may be instructive to  write down explicitly the K\"ahler and super-potentials in terms of $H_u$ and $H_d$:
\bea
\label{Kand W3}
K&=&\kappa_u|H_u|^2 + \kappa_d |H_d|^2 - \frac{1}{2} \left(H_uH_d- (H_u H_d)^{\dag}\right)^2 \non \\&&+ |X|^2,\\
\label{Kand W4}
W&=&\lambda X H_uH_d,
\eea
with $\kappa_u + \kappa_d = \kappa$. We note that the form of the superpotential (\ref{Kand W4}) is equivalent to the part of the interactions in NMSSM.

The scalar potential in supergravity is given by
\bea
V&=& e^K \left(D_i W K^{i\bar{j}} (D_j W)^* - 3 |W|^2 \right).
\label{sugraV}
\eea
Since we have imposed the $Z_4$ symmetry, $\phi^2 -\phi^{\dag 2} \approx 0$ along the inflationary trajectory.
The relevant Lagrangian for the inflation is then given by
\bea
\label{lagrangian}
{\cal L}& = &  \left(\kappa + 2^2 |\phi|^{2} \right) \partial^\mu \phi^\dag \partial_\mu \phi - V(\phi),\\
V(\phi)
&\approx&
e^{\kappa |\phi|^2}\, \lambda^2 |\phi|^{4}.
\label{V}
\eea
Since we explicitly break the shift symmetry (\ref{sym}) by the
$\kappa$ term, there appears a non-vanishing exponential
prefactor. However, for $|\phi| < 1/\sqrt{\kappa}$, the exponential
prefactor is close to unity, and therefore can be dropped. Note that the
inflaton does slow-roll even if the exponential pre-factor gives a main
contribution to the tilt of the potential, as long as $\kappa$ is much 
smaller than unity. Except for the exponential factor, one can see
 that $\xi$ and $\lambda_h$ in Eq.~(\ref{SM}) are related to $\kappa$ and $\lambda$ as 
$\xi = 4/\kappa^2$ and $\lambda_h = \lambda^2/\kappa^2$.

\begin{table}[t!]
\begin{center}
\begin{tabular}{c||c|c|c|c}
                   &\,   $H_u$ \,  & \,  $H_d$  \, &\,  $\phi^2$  \,& \, $X$ \\ \hline
 U(1)$_R$&$0$&$0$&$0$&$2$ \\ \hline
 $Z_4$ & $1$  & $1$ & $2$& $2$ 
\end{tabular}
\end{center}
\caption{The charge assignment of $\phi$ and $X$ in the $\phi^2$ chaotic inflation.}
\label{charge}
\end{table}

For $1 < |\phi| \ll \kappa^{-1/2}$, the Lagrangian can be approximated
by
\bea
{\cal L}&\approx&  2^2 |\phi|^{2} \partial^\mu \phi^\dag \partial_\mu \phi - \lambda^2 |\phi|^{4},\\
&=& \partial^\mu \phih^\dag \partial_\mu \phih - \lambda^2 |\phih|^{2},
\eea
where we have defined $\phih \equiv \phi^2$. The inflationary trajectory is given by $\phi^2 = \phi^{\dag 2}$,
and so, the imaginary component of $\phih$ vanishes. Let us rewrite the inflaton as
\beq \phih \;= \; \frac{\varphi}{\sqrt{2}}, 
\eeq 
where
$\varphi$ is a real scalar.  The Lagrangian for the canonically normalized inflaton is
therefore given by
\beq
\label{V_can}
{\cal L}\; \approx \; \frac{1}{2} \partial^\mu \varphi \partial_\mu \varphi - \frac{\lambda^2}{2} \varphi^2,
\eeq
for $1 < \varphi \ll \kappa^{-1}$. Thus, thanks to the shift
symmetry, the inflaton $\varphi$ can take a value greater than the
Planck scale, and the chaotic inflation takes place.

The inflaton field durning inflation is related to the e-folding number $N$ as
\beq
\varphi_N \;\simeq\;\sqrt{4N},
\eeq
and the inflation ends at $\varphi \approx 1$. The power
spectrum of the density perturbation is given by
\beq
\Delta_{\cal R}^2 \;\simeq\; \frac{V^{3}}{12 \pi^2 V^{\prime 2}} = (2.43 \pm 0.11)\times 10^{-9},
\eeq
where we have used in the second equality the WMAP
result~\cite{Komatsu:2010fb}. The coupling $\lambda$ is
therefore determined as
\beq
 \lambda \;\simeq\;  8\times 10^{-6}\lrfp{N}{50}{-\frac{1}{2}} \simeq\; 2 \times \GEV{13} \lrfp{N}{50}{-\frac{1}{2}}.
\eeq

In order for the inflation driven by (\ref{V_can}) to last for $N$
e-foldings, the following inequality must be met;
\beq
\varphi_N \;\lesssim\;\kappa^{-1} ~~\Longleftrightarrow~~~ \kappa \;\lesssim\; 0.07 \,\lrfp{N}{50}{\frac{1}{2}}.
\eeq
The spectral index $n_s$ and the tensor-to-scalar ratio $r$ are respectively given by
\bea
n_s &=& 1- \frac{2}{N},\\
r &=& \frac{8}{N}.
\eea
For $N=50 \sim 60$, they vary as $n_s = 0.96 \sim 0.967$ and $r=0.13 \sim 0.16$.

The inflation ends when the slow-roll condition is violated at
$\varphi \sim 1$, and the inflaton starts to oscillate about the
origin. The dynamics of the inflaton is then described by a complex scalar field $\phih$ rather than
the real scalar $\varphi$. As the amplitude of the inflaton decreases, the
$\kappa$ term becomes more important. For $|\phi| <
(\kappa/4)^{1/2}$, the Lagrangian becomes
\beq
{\cal L}\;\approx\;  \partial^\mu \phit^\dag \partial_\mu \phit- \frac{\lambda^2}{\kappa^2} |\phit|^{4},
\label{L_at_lowE}
\eeq
where we have defined a canonically normalized field at low scales,
$\phit \equiv \sqrt{\kappa} \phi$.  Note that the power of the scalar potential
changes from $2$ to $4$ after inflation. When the amplitude becomes of the order of the weak scale,
the description by the D-flat direction $H_uH_d$ is no longer valid, and we should consider
the dynamics of $H_u$ and $H_d$ separately as usual. In the end, they should develop vacuum
expectation values (VEVs), leading to the electroweak phase transition.

In order to have
a successful electroweak phase transition, the $\mu$-term with a right magnitude must be generated. 
We may add small explicit breaking of the discrete symmetry to produce a tadpole of $X$, which
makes $X$ to develop a VEV, generating the $\mu$-term.
Alternatively, we may  identify the $X$ field as the singlet field in the NMSSM, in which the
superpotential takes the following form,
\beq
W\;=\; \lambda X H_uH_d + y \frac{X^3}{3},
\eeq
where $y$ is a coupling constant. 
Note that the presence of $X^3$ in the superpotential does not destabilize the
inflation dynamics. In order to have a chaotic inflation in NMSSM, we need to consider a different shift symmetry. 
Instead of the $Z_4$ symmetry, let us assign $Z_3$ symmetry
on $X$ and the Higgs field.\footnote{If the $Z_3$ is exact, domain walls will be produced. To avoid the domain-wall problem
we need to introduce a small $Z_3$ breaking.} See Table~\ref{charge2}. The simplest shift symmetry consistent 
with the $Z_3$ symmetry is given by\footnote{
We can also consider a shift symmetry $\phi^{6 \ell} \rightarrow \phi^{6 \ell} + \alpha$ with $\ell = 2,3,\cdots$.
The potential would be proportional to $\varphi^{2/3\ell}$ where $\varphi \sim \phi^{6 \ell}$.
}
\beq
\phi^6 \rightarrow \phi^6 + \alpha.
\label{sym2}
\eeq
Along the same line, we can realize a chaotic inflation with the Higgs fields $H_uH_d$ as the inflaton.
The K\"ahler potential is given by
\beq
K\;=\; c_1 \,(\phi^6 - \phi^{\dag 6}) -\, \frac{1}{2}(\phi^6 - \phi^{\dag 6})^2 + \cdots,
\label{Kahler2}
\eeq
where $c_1$ is in general non-zero.
The potential is given by
\beq
V(\varphi) \;\approx\; \lambda^2 \lrfp{ \varphi}{\sqrt{2}}{\frac{2}{3}},
\eeq
where $\varphi = \sqrt{2}(\phi^6- c_1/2)$ is the canonically normalized field. The spectral index and the tensor-to-scalar
ratio, respectively, are $n_s = 0.973 \sim 0.978$ and $r=0.044 \sim 0.053$ for $N=50 \sim 60$.
The WMAP normalization gives $\lambda \simeq 2 \times 10^{-5}$.

Note that, while $\lambda \sim 10^{-5}$ is determined by the WMAP normalization, 
the low-energy effective coupling between the singlet $X$ and the Higgs is given by
\beq
\hat{\lambda}\;=\;\frac{\lambda}{\sqrt{\kappa_u \kappa_d}}.
\eeq
So, if $\lambda \sim \kappa_u \sim \kappa_d$, the effective coupling can be $O(0.1)$.
Similarly, the SM Yukawa interactions break the shift symmetry:
\beq
W_{MSSM} \;=\; y_u Q {\bar u} H_u + y_d Q {\bar d} H_d + y_e L {\bar e} H_d,
\eeq
where $|y_{u,d,e}| \ll 1$ are the Yukawa couplings, and we suppressed the generations. 
The physical Yukawa couplings at the low energy are similarly scaled as
\beq
y_u^{(phys)} = \frac{y_u}{\sqrt{\kappa_u}},~~y_d^{(phys)} = \frac{y_d}{\sqrt{\kappa_d}},~~y_e^{(phys)} = \frac{y_e}{\sqrt{\kappa_d}},
\eeq
Therefore the top Yukawa coupling can be close to $1$, if $y_u \sim \sqrt{\kappa_u} = O(10^{-3})$.  
The coefficients of the 
breaking terms are suppressed by a factor of $O(10^{-3})$ wherever either $H_u$ or $H_d$ appears: $y_u^{(top)} \sim 10^{-3}$,
and $\lambda \sim \kappa_u \sim 10^{-6} - 10^{-5}$. This structure might be related with the UV theory behind the shift symmetry.

We emphasize here that the presence of $X$ is essential for constructing a chaotic inflation model in supergravity.
It is stabilized at the origin and its dynamics is not relevant for the inflation, and so, $X$ may be considered as a spectator field.
Interestingly, however, in the Higgs chaotic inflation model, the same $X$ plays an important role in low-energy phenomenology. 
For instance, in the NMSSM, the fermionic superpartner of $X$ can be dark matter.

So far we have considered the possibility that the Higgs fields play the role of the inflation. It is straightforward to apply
the above idea to the other flat directions in MSSM. In this case we need to adopt a flat direction which is lifted by
the superpotential of the form $W \sim X \phi^m$~\cite{Gherghetta:1995dv}. (For instance the $LLe$ direction could be
lifted by $W = H_u L LL e$, and $X$ is identified with $H_u$).
In particular, if the flat direction has a non-zero baryon/lepton number, 
the baryon/lepton numbers would be explicitly
violated by the interactions in the K\"ahler potential, and so, the baryogenesis a la Affleck-Dine~\cite{Affleck:1984fy,Dine:1995kz} 
is possible~\cite{Takahashi:2010ky,Nakayama:2010kt}. There would be no baryonic isocurvature perturbation~\cite{Kasuya:2008xp} 
because the degree of the freedom orthogonal to the inflaton is heavy during inflation.

Let us briefly mention the reheating in the Higgs inflation model. The SM particles are naturally created
by the inflaton decay in the Higgs inflation, but the process could be complicated by the non-perturbative decay.
In the SM Higgs inflation and the NMSSM Higgs inflation with the $Z_4$ symmetry, the inflaton passes near the
origin after inflation, and so, the preheating is likely to occur~\cite{Bezrukov:2008ut,GarciaBellido:2008ab}. 
On the other hand, in the last example, the inflaton acquires a non-zero angular momentum due to the
non-zero $c_1$. Then the preheating may not be efficient. If the non-perturbative decay is efficient in the former case,
the resultant reheating temperature would be very high, and too many gravitinos may be produced from thermal scattering, while
the non-thermal gravitino production~\cite{Kawasaki:2006gs,Endo:2007ih} is generically suppressed in the Higgs inflation. 

If the Higgs chaotic inflation is realized
in nature, we will be able to study the properties of the inflaton, namely the Higgs fields,
at the collider experiments as well as the CMB observation.

\begin{table}[t!]
\begin{center}
\begin{tabular}{c||c|c|c|c}
                   &\,   $H_u$ \,  & \,  $H_d$  \, &\,  $\phi^2$  \,& \, $X$ \\ \hline
 $Z_3$ & $1$  & $1$ & $2$& $1$ 
\end{tabular}
\end{center}
\caption{The charge assignment of $\phi$ and $X$ in the $\phi^{2/3}$ chaotic inflation.}
\label{charge2}
\end{table}

%
\begin{acknowledgements}
  
  This work was supported by the Grant-in-Aid for Scientific Research
  on Innovative Areas (No. 21111006) [KN and FT] and Scientific
  Research (A) (No. 22244030) [FT], and JSPS Grant-in-Aid for Young
  Scientists (B) (No. 21740160) [FT].  This work was supported by
  World Premier International Center Initiative (WPI Program), MEXT,
  Japan.

\end{acknowledgements}




\begin{thebibliography}{}

\bibitem{Komatsu:2010fb}
  E.~Komatsu {\it et al.},
  [arXiv:1001.4538 [astro-ph.CO]].
  
\bibitem{:2006uk}
    [Planck Collaboration],
  arXiv:astro-ph/0604069.
  
\bibitem{Murayama:1992ua}
  H.~Murayama, H.~Suzuki, T.~Yanagida {\it et al.},
  Phys.\ Rev.\ Lett.\  {\bf 70}, 1912-1915 (1993).
 

\bibitem{Kasuya:2003iv}
  S.~Kasuya, T.~Moroi, F.~Takahashi,
  Phys.\ Lett.\  {\bf B593}, 33-41 (2004).
  [hep-ph/0312094].
  
\bibitem{Allahverdi:2006iq}
  R.~Allahverdi, K.~Enqvist, J.~Garcia-Bellido {\it et al.},
  Phys.\ Rev.\ Lett.\  {\bf 97}, 191304 (2006).
  [hep-ph/0605035].  
  
\bibitem{Bezrukov:2007ep}
  F.~L.~Bezrukov, M.~Shaposhnikov,
  Phys.\ Lett.\  {\bf B659}, 703-706 (2008).
  [arXiv:0710.3755 [hep-th]].



\bibitem{Salopek:1988qh}
  D.~S.~Salopek, J.~R.~Bond, J.~M.~Bardeen,
  Phys.\ Rev.\  {\bf D40}, 1753 (1989);
  T.~Futamase, K.~-i.~Maeda,
  Phys.\ Rev.\  {\bf D39}, 399-404 (1989);
  B.~L.~Spokoiny,
  Phys.\ Lett.\  {\bf B147}, 39-43 (1984);
  R.~Fakir, W.~G.~Unruh,
  Phys.\ Rev.\  {\bf D41}, 1783-1791 (1990);
  E.~Komatsu, T.~Futamase,
  Phys.\ Rev.\  {\bf D58}, 023004 (1998).
  [astro-ph/9711340].
 


\bibitem{Einhorn:2009bh}
  M.~B.~Einhorn and D.~R.~T.~Jones,
  JHEP {\bf 1003}, 026 (2010)
  [arXiv:0912.2718 [hep-ph]].




\bibitem{Lee:2010hj}
  H.~M.~Lee,
  JCAP {\bf 1008}, 003 (2010).
  [arXiv:1005.2735 [hep-ph]].

\bibitem{Ferrara:2010yw}
  S.~Ferrara, R.~Kallosh, A.~Linde {\it et al.},
  Phys.\ Rev.\  {\bf D82}, 045003 (2010).
  [arXiv:1004.0712 [hep-th]].

\bibitem{Ferrara:2010in}
  S.~Ferrara, R.~Kallosh, A.~Linde {\it et al.},
  [arXiv:1008.2942 [hep-th]].
  
\bibitem{Kallosh:2010ug}
  R.~Kallosh, A.~Linde,
  [arXiv:1008.3375 [hep-th]].

\bibitem{BenDayan:2010yz}
  I.~Ben-Dayan and M.~B.~Einhorn,
  arXiv:1009.2276 [hep-ph].


\bibitem{Takahashi:2010ky}
  F.~Takahashi,
  Phys.\ Lett.\  B {\bf 693}, 140 (2010)
  [arXiv:1006.2801 [hep-ph]].

    
     

\bibitem{Nakayama:2010kt}
  K.~Nakayama and F.~Takahashi,
  arXiv:1008.2956 [hep-ph].



\bibitem{Dimopoulos:2003iy}
  S.~Dimopoulos, S.~D.~Thomas,
  Phys.\ Lett.\  {\bf B573}, 13-19 (2003).
  [hep-th/0307004].

\bibitem{Izawa:2007qa}
  K.~-I.~Izawa, Y.~Shinbara,
  [arXiv:0710.1141 [hep-ph]].
  
\bibitem{Linde:1983gd}
  A.~D.~Linde,
  Phys.\ Lett.\ B {\bf 129}, 177 (1983).


\bibitem{Lyth:1996im}
  D.~H.~Lyth,
  Phys.\ Rev.\ Lett.\  {\bf 78}, 1861 (1997).
  
\bibitem{Germani:2010gm}
  C.~Germani and A.~Kehagias,
  Phys.\ Rev.\ Lett.\  {\bf 105}, 011302 (2010)
  [arXiv:1003.2635 [hep-ph]].
  
  
\bibitem{NT}
K.~Nakayama and F.~Takahashi, in preparation.   

\bibitem{Kawasaki:2000yn}
  M.~Kawasaki, M.~Yamaguchi, T.~Yanagida,
  Phys.\ Rev.\ Lett.\  {\bf 85}, 3572-3575 (2000).

\bibitem{Gherghetta:1995dv}
  T.~Gherghetta, C.~F.~Kolda, S.~P.~Martin,
  Nucl.\ Phys.\  {\bf B468}, 37-58 (1996).
  [hep-ph/9510370].  
  
\bibitem{Affleck:1984fy}
  I.~Affleck and M.~Dine,
  Nucl.\ Phys.\  B {\bf 249}, 361 (1985).
  
  
\bibitem{Dine:1995kz}
  M.~Dine, L.~Randall, S.~D.~Thomas,
  Nucl.\ Phys.\  {\bf B458}, 291-326 (1996).
  [hep-ph/9507453].

\bibitem{Kasuya:2008xp}
  S.~Kasuya, M.~Kawasaki and F.~Takahashi,
  JCAP {\bf 0810}, 017 (2008)
  [arXiv:0805.4245 [hep-ph]].

  
\bibitem{Bezrukov:2008ut}
  F.~Bezrukov, D.~Gorbunov and M.~Shaposhnikov,
  JCAP {\bf 0906}, 029 (2009)
  [arXiv:0812.3622 [hep-ph]].

\bibitem{GarciaBellido:2008ab}
  J.~Garcia-Bellido, D.~G.~Figueroa and J.~Rubio,
  Phys.\ Rev.\  D {\bf 79}, 063531 (2009)
  [arXiv:0812.4624 [hep-ph]].

  
\bibitem{Kawasaki:2006gs}
  M.~Kawasaki, F.~Takahashi, T.~T.~Yanagida,
  Phys.\ Lett.\  {\bf B638}, 8 (2006);
  Phys.\ Rev.\  D {\bf 74}, 043519 (2006).
  
\bibitem{Endo:2007ih}
  M.~Endo, F.~Takahashi, T.~T.~Yanagida,
  Phys.\ Lett.\  {\bf B658}, 236 (2008);
  Phys.\ Rev.\  {\bf D76}, 083509 (2007).


\end{thebibliography}
\end{document}